\documentclass[10pt, conference, letterpaper]{IEEEtran}

\usepackage{url,xspace,graphicx}
\usepackage{subcaption}
\usepackage{units}
\usepackage{float}
\usepackage{listings,mdframed}
\usepackage{array}
\usepackage{verbatim}
\usepackage{color}

\newlength{\grafflecm}
\usepackage{pslatex}
\usepackage{clrscode}
\usepackage{theorem}
\usepackage{amsmath}
\usepackage{amsfonts}
\usepackage{amssymb}

\newcommand*{\TechRep}{}%

\IEEEoverridecommandlockouts
\begin{document}
\interfootnotelinepenalty=10000
%%%%%%%%%%%%%%%%%%%%%%%%%%%%%%%%%%%%%%%%

\title{
\ifdefined\TechRep
FM-Delta: Fault Management Packet Compression 
\newline
\newline
				\large Technical Report\textsuperscript{\ensuremath\diamond}
				\thanks{\textsuperscript{\ensuremath\diamond}This technical report is an extended version of~\cite{Fmdelta}, which was accepted to the IFIP/IEEE International Symposium on Integrated Network Management, IM 2017.}, January 2017
\else
FM-Delta: Fault Management Packet Compression
\fi
}

\author{Tal Mizrahi\textsuperscript{\ensuremath*}\textsuperscript{\ensuremath\dagger}, Yoram Revah\textsuperscript{\ensuremath*}, Yehonathan Refael Kalim\textsuperscript{\ensuremath\dagger}, Elad Kapuza\textsuperscript{\ensuremath\dagger}, Yuval Cassuto\textsuperscript{\ensuremath\dagger}
\\ \textsuperscript{\ensuremath*}Marvell Semiconductors, \textsuperscript{\ensuremath\dagger}Technion
\\ talmi@marvell.com, revahyo@gmail.com, \{srefaelk@campus, eladkap@campus, ycassuto@ee\}.technion.ac.il
}

\maketitle
\thispagestyle{empty}

\begin{abstract}
%\begin{sloppypar}
Fault Management (FM) is a cardinal feature in communication networks. One of the most common FM approaches is to use periodic keepalive messages. Hence, switches and routers are required to transmit a large number of FM messages periodically, requiring a hardware-based packet generator that periodically transmits a set of messages that are stored in an expensive on-chip memory. 
With the rapid growth of carrier networks, and as 5G technologies emerge, the number of users and the traffic rates are expected to significantly increase over the next few years. Consequently, we expect the on-chip memories used for FM to become a costly component in switch and router chips. 
We introduce a novel approach in which FM messages are stored in compressed form in the on-chip memory, allowing to significantly reduce the memory size. We present FM-Delta, a simple hardware-friendly delta encoding algorithm that allows FM messages to be compressed by a factor of 2.6. 
We show that this compression ratio is very close to the results of the zlib compression library, which requires much higher implementation complexity.
%\end{sloppypar}
\end{abstract}

%\keywords{Fault management, OAM, compression.}

\section{Introduction}
\label{IntroSec}
\subsection{Background}
Network devices, such as switches and routers, are often required to transmit control-plane messages. The ability to generate and transmit messages has been recognized as an essential building block in network devices, not only in conventional networks, but also in Software-Defined Networks (e.g.,~\cite{ONFPIF}). Specifically, the ability to generate packets is important in the context of Fault Management (FM).

Fault detection is essential in large-scale networks, enabling fast recovery and effective troubleshooting; it is one of the key components in Operations, Administration, and Maintenance (OAM)~\cite{mizrahi2014overview}. FM is typically implemented as a combination of proactive and reactive mechanisms for detecting failures. Some of the most commonly deployed FM protocols (e.g., ~\cite{IEEE8021ag,Y1731,RFC5880}) are implemented using periodic keepalive messages; a fault is detected when no keepalive messages are received from a given source for a given period of time. 

\subsection{FM Scaling}
The scaling problem of storing FM messages is a real-life problem that some of the authors of this paper encountered while designing packet processor silicons. A switch or a router that runs an FM protocol periodically transmits FM messages. The rate of FM messages varies from 1 packet every ten minutes to 300 packets per second per service~\cite{IEEE8021ag,Y1731}. Typical carrier Ethernet devices support tens of thousands of services (e.g., 16k in~\cite{EricssonSPO1400} and 64k in~\cite{Huawei}). Due to these large scales, typical devices do not provide FM support at the highest rate for all services simultaneously. As~5G technologies evolve, network devices will be required to support a larger number of services at a higher bandwidth. Thus, fault detection will be expected with a low detection time, implying a high rate of FM messages. Hence, we expect these scales to continuously increase in the next few years.

\textbf{Example.}
The rate of traffic that a network device is expected to generate for 64k services, assuming the FM message length is roughly 100B~\cite{IEEE8021ag}, and assuming 300 packets per second per service~\cite{IEEE8021ag}, is $64k \times 300 \times 100B \cong 16~Gbps$. Such high traffic rates cannot be handled by the device's software layer, and thus must be implemented in the device's hardware. A typical implementation (e.g.,~\cite{mizrahi2011oam}) uses a hardware engine and an on-chip memory; the engine periodically reads the messages stored in the memory and transmits them to the network. In this case the required memory space for 64k services is $64k \times 100B \cong 6.4~MB$. This is a very significant size for an on-chip memory, consuming an expensive area of roughly 18 $mm^2$ in a 28nm process. As a point of reference, the entire on-chip packet memory of a typical switch is on the order of a few megabytes, e.g., 12 MB in~\cite{pica8}.

This example illustrates that as carrier and mobile backhaul networks evolve, the significant size of the FM on-chip memory may become infeasible.

\subsection{FM Packet Compression in a Nutshell}
We argue that if FM packets are compressed when stored in the memory, then the expensive on-chip memory can be significantly reduced. 

In our approach FM packets are compressed offline, and stored in compressed form in an on-chip memory. Each packet is decompressed by the hardware packet generation engine before it is transmitted to the network. Notably, packets are in compressed form only when stored in memory, and are decompressed before transmission, making the compression transparent to the network.

\textbf{Why do we focus on FM messages?} The solution we propose relies on three properties that are unique to the problem at hand: (i)~the entropy of the FM packets stored in the memory is low, as they share common properties, (ii)~the FM packet memory is accessed sequentially, since all packets are sent periodically, and (iii)~we can determine the \emph{order} of the FM packets in the memory.\footnote{In the FM-Delta system we present we have full control of the order in which packets are stored in the memory, and thus the order in which packets are accessed and decompressed.} These three properties significantly improve the effectiveness of the compression.

\subsection{Related Work}
Data compression has been analyzed in the context of network switches and routers, e.g.,~\cite{rottenstreich2014compressing,retvari2013compressing}. Specifically, data-plane packet compression has been widely discussed and analyzed in the literature, e.g.,~\cite{tye2003review,gutwin2006improving}. For example, packet compression is widely used in the Hypertext Transfer Protocol (HTTP)~\cite{rfc2616}. It has been shown that HTTP compression can be very effective~\cite{websiteOptimization}, in some cases reducing the data to 20\% of its original size. However, 
%in the general case packet compression is not guaranteed to be effective. It 
it has been shown~\cite{tye2003review} that packet compression is ineffective for random internet traffic, as this traffic typically has high entropy, and is often already compressed or encrypted. In this paper we focus on control-plane packet compression, and present a use case in which packet compression is highly effective.

%The Lempel-Ziv (LZ) algorithm~\cite{ziv1978compression} is a one of the most commonly used compression algorithms. One of its most well-known open source implementations is zlib~\cite{zlib}. In this paper we evaluate the effectiveness of the LZ algorithm on FM packets, and compare it to a delta encoding approach.

\subsection{Contributions}
The contributions of this paper are as follows:

\begin{itemize}
	\item We present a novel approach that uses packet compression to reduce the size of the on-chip memory used in Fault Management (FM) message transmission. 
	\item We introduce FM-Delta, a simple delta encoding algorithm that can easily be implemented in hardware, and present a high-level design of a system that uses FM-Delta. 
  \item We evaluated the algorithm over a synthetically generated FM packet database, and show that it offers a compression ratio that is comparable to state-of-the-art compression algorithms, such as Lempel-Ziv (LZ).\footnote{It is important to note that known LZ implementations (such as zlib~\cite{zlib}) are unlikely to be realizable in the low-level hardware setup of interest here, due to their significant processing and memory requirements.} The FM packet database that we used in our experiments is publicly available~\cite{datasets}.
\end{itemize}

\widowpenalties 1 10000

%\ifdefined\CutSpace\vspace{-2mm}\fi
\section{A System Design}
%\ifdefined\CutSpace\vspace{-1mm}\fi
\subsection{Overview}
\begin{figure}[tbp]
	\ifdefined\CutSpace\vspace{-2mm}\fi
	\centering
  \includegraphics[height=4.5\grafflecm]{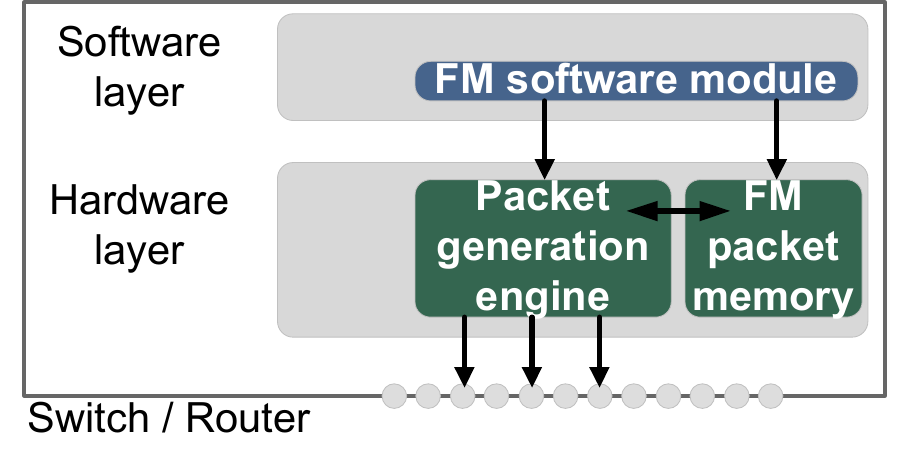}
	\captionsetup{justification=raggedright}
	%\vspace{-3mm}
  \caption{FM-Delta: a system design.}
  \label{fig:SystemDesign}
	\ifdefined\CutSpace\vspace{-2mm}\fi
\end{figure}

Fig.~\ref{fig:SystemDesign} illustrates the design of a system that uses FM-Delta to periodically generate FM packets. The system consists of two main components:

\begin{itemize}
	\item Software module -- responsible for compressing the FM packets and storing them in the on-chip memory of the packet generation engine. The compression procedure is performed offline. When an FM packet needs to be added to the FM packet database, or removed from it, the software layer invokes an \emph{insert} or a \emph{remove} operation, instructing the packet generation engine to perform an incremental update in the FM database.
	\item Packet generation engine -- this hardware module sequentially reads the packets from the on-chip FM packet memory, decompresses them in real-time, and transmits them to the network.
\end{itemize}

\subsection{Compression Algorithm}
We introduce FM-Delta, a simple delta encoding algorithm that we designed and implemented.
%In the current work we analyze data compression using two compression schemes: (i) Lempel-Ziv compression, using the zlib~\cite{zlib} library, and (ii) FM-Delta; a simple Delta encoding algorithm that we designed and implemented.

\textbf{FM-Delta.}
We present an outline of the FM-Delta algorithm. Given a sequence of N uncompressed packets, the compressed packets are represented as follows.

\begin{figure}[htbp]
	\ifdefined\CutSpace\vspace{-2mm}\fi
	\centering
  \includegraphics[height=.8\grafflecm]{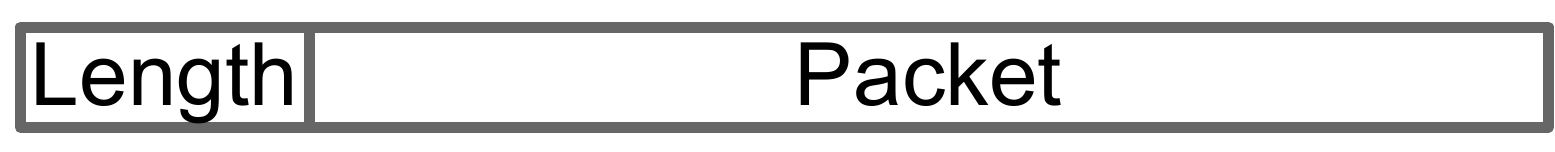}
	\captionsetup{justification=raggedright}
	\ifdefined\CutSpace\vspace{-1.5mm}\fi
  \caption{FM-Delta: the first packet is in uncompressed form.}
  \label{fig:First}
\end{figure}

The first packet is represented in uncompressed form, as shown in Fig.~\ref{fig:First}.

\begin{figure}[htbp]
	\ifdefined\CutSpace\vspace{-2mm}\fi
	\centering
  \includegraphics[height=.8\grafflecm]{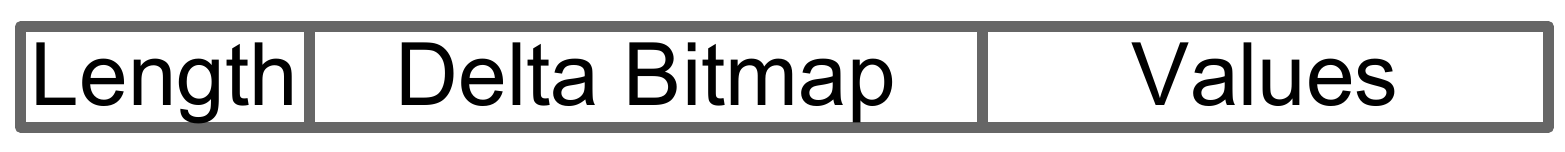}
	\captionsetup{justification=raggedright}
	\ifdefined\CutSpace\vspace{-1.5mm}\fi
  \caption{FM-Delta: structure of the $i^{th}$ compressed packet.}
  \label{fig:Compressed}
\end{figure}

The $i^{th}$ packet, for $i\geq 2$, is compressed, as shown in Fig.~\ref{fig:Compressed}.

%\begin{comment}
\begin{figure}[!b]
\ifdefined\CutSpace\vspace{-2mm}\fi
\hrule
\ifdefined\CutSpace\vspace{-2mm}\fi
  \begin{codebox}
    \Procname{$\proc{Decompress}(N)$}
		\li Read packet $P_1$
		\li $L_1 \leftarrow$ Length of $P_1$
		\li $U_1 \leftarrow$ uncompressed packet $P_1$
		\li for $i=2$ to $N$
		\li \ \ Read compressed packet $P_i$
		\li \ \ $D \leftarrow$ Delta Bitmap of $P_i$
		\li \ \ $V \leftarrow$ Values of $P_i$
		\li \ \ $m \leftarrow 0$
		\li \ \ for $j=0$ to $min(L_i, L_{i-1})-1$
		\li \ \ \ \ if $D[j]=1$
		\li \ \ \ \ \ \ $U_i[j]\leftarrow U_{i-1}[j]$
		\li \ \ \ \ else
		\li \ \ \ \ \ \ $U_i[j]\leftarrow V[m]$
		\li \ \ \ \ \ \ $m\leftarrow m+1$
		
  \end{codebox}
	\ifdefined\CutSpace\vspace{-2mm}\fi
  \hrule
  \caption{Decompression algorithm for $N$ packets. The $i^{th}$ compressed packet is denoted by $P_i$, and the $i^{th}$ uncompressed packet by $U_i$.}
  \label{fig:DecompAlg}

\end{figure}

%\end{comment}

The compressed packet consists of the following fields (Fig.~\ref{fig:Compressed}): 
\begin{itemize}
	\item \textbf{Length} -- represents the length in bytes of the original (uncompressed) packet. 
	\item \textbf{Delta Bitmap} -- indicates the differences between packet $i$ and packet $i-1$. Every packet is divided into equal-sized words, and each bit in the bitmap indicates whether the respective word in the current packet is equal to the corresponding word in the previous packet. The word size is a parameter of the algorithm. Section~\ref{EvaluationSec} discusses the effectiveness of the algorithm with various word sizes. 
	\item \textbf{Values} -- the values of the words that differ from the previous packet.
\end{itemize}

%The compression procedure can be performed offline, storing the compressed packets in the on-chip memory. The decompression is performed in real-time, allowing each packet to be decompressed immediately before it is transmitted. 
The decompression algorithm is presented in Fig.~\ref{fig:DecompAlg}.

Common (de)compression algorithms, such as Lempel-Ziv~\cite{ziv1978compression} are not hardware-friendly, as they require a complex iterative indirection over a sliding window. In FM-Delta, every packet is decompressed by comparing it to the previous (decompressed) packet using the Delta Bitmap. This simple comparison does not require an iterative procedure, and thus all the words of the packet can be compared (decompressed) in parallel, making FM-Delta a silicon-friendly algorithm.

\subsection{Insertion and Removal}
As described above, the packet database is compressed offline, and accessed by the chip in real-time. An essential question is how the packet database is updated, i.e., how a new packet can be added to the database, or how a packet can be removed from it.

One approach that allows simple insertion and removal is by using a linked list. Each compressed packet in the memory is followed by a pointer that points to the location of the next packet in memory. The linked list approach is simple, but requires some memory overhead for the next pointers, and also may be inefficient due to fragmentation.
%Thus inserting a new packet into location $j$ requires atomically performing two actions: recompressing packet $j$, and updating the pointer of packet $j-1$.  

\begin{figure}[tbp]
	\ifdefined\CutSpace\vspace{-2mm}\fi
	\centering
  \includegraphics[height=4.7\grafflecm]{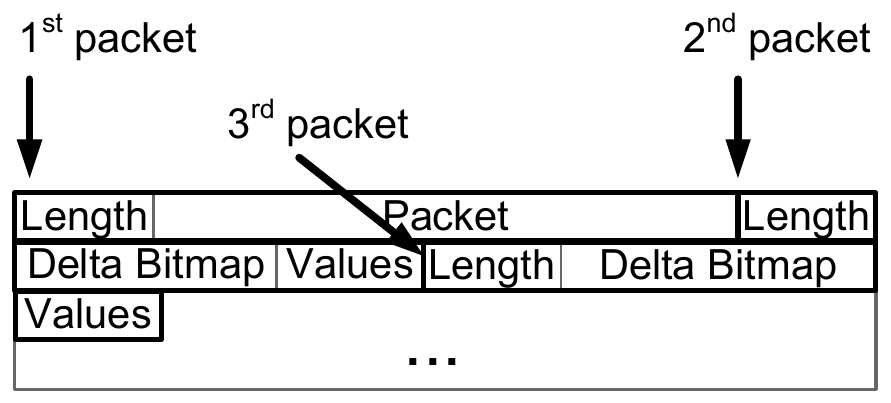}
	\captionsetup{justification=raggedright}
	%\vspace{-3mm}
  \caption{FM-Delta: the compressed packets in the memory.}
  \label{fig:Memory}
\end{figure}

We suggest a more efficient approach that allows for simple insertion and removal of packets when using FM-Delta. In this approach (Fig.~\ref{fig:Memory}) packets are stored contiguously in the memory. The packet generation engine proceeds by a read-decompress-write procedure; every packet is read from memory, decompressed, and then written back to the memory. There is a cache that stores the previous packet, which enables the engine to decompress the current packet using FM-Delta.

\ifdefined\TechRep

\vspace{3mm}
\textbf{Removal} 
\vspace{1mm}

Assume that we have $N$ compressed packets in the memory, denoted by $P_1, P_2, \ldots, P_N$, and we want to remove a packet $P_k$ from the database. Removing the $k^{th}$ packet from the database includes two operations that need to take place atomically: (i)~removing the $k^{th}$ packet from the memory, and (ii)~updating the $(k+1)^{th}$ delta-encoded packet. The latter is required since after removing packet $P_k$, the compressed $(k+1)^{th}$ packet is encoded with respect to packet $P_{k-1}$. The two operations must be performed atomically to prevent inconsistent reading or decompression.

In order to perform the two operations above atomically, we assume that there is a software layer that triggers the removal operation, and that the removal itself is performed by the packet generation engine. When the software layer triggers the operation, it also provides $k$, the index of the packet to be removed, and $P'_{k+1}$, the newly compressed $(k+1)^{th}$ packet.

%Assume we have $N$ packets in the memory, denoted by $P_1, P_2, \ldots, P_N$, and we want to remove a packet $P_k$ from the database. 
%A removal event should be triggered by the software layer; the software layer invokes a removal operation, which includes $k$, the index of the packet to be removed, and $P'_{k+1}$, 

Once the packet generation engine receives the removal request it: 

\begin{enumerate}
	\item Performs the conventional read-decompress-write procedure until it reaches packet $P_{k-1}$.
	\item Reads and decompresses packets $P_k$ and $P_{k+1}$, but does not write them back to the memory.
	\item Writes the newly compressed packet, $P'_{k+1}$ immediately after packet $P_{k-1}$.
	\item Continues the read-decompress-write procedure on packets $P_{k+2}, \ldots, P_N$, so that each packet is written after the previously written packet, thus eliminating the gap created by removing $P_k$.
\end{enumerate}
  
\vspace{3mm}
\textbf{Insertion} 
\vspace{1mm}

Assume we have $N$ packet in the memory, and we want to insert a new compressed packet, $P$, before packet $P_k$. As in the removal procedure, two operations need to occur atomically: (i)~inserting packet $P$, and (ii)~updating packet $P_k$ to a newly encoded $P'_k$. Again, we assume that there is a software layer that provides the location $k$, the packet that needs to be inserted $P$, and the newly encoded $P'_k$.

Upon receiving an insertion request, the packet generation engine:

\begin{enumerate}
	\item Performs the conventional read-decompress-write procedure until it reaches packet $P_{k-1}$.
	\item Reads and decompresses packet $P_k$, but does not write it back to the memory.
	\item Writes the newly inserted (compressed) packet $P$ immediately after $P_{k-1}$.
	\item For packets $P_{k+1}, \ldots, P_N$, the engine proceeds by reading packet $P_j$, decompressing it, and then writing packet $P_{j-1}$ after the previously written packet.
\end{enumerate}

Note that in steps 2-4 above, the packet generation engine takes care not to write over packets that have not yet been read. For example, if the compressed packet $P_j$ is slightly shorter than $P_{j-1}$, the engine writes $P_{j-1}$ only after having read $P_{j+1}$. Thus, the packet generation engine maintains a small cache that stores the currently read packets.

\else
The packet ordering, insertion, and removal are described in further detail in the extended version of this paper~\cite{FMdeltaTR}.
\fi

\subsection{Random Access to FM Packets}
In our approach we assumed that the FM packet memory is always accessed sequentially. However, in some cases the FM application may require an \emph{urgent} packet $P_k$ to be sent immediately. The delta encoding scheme implies that in order to access the $k^{th}$ packet the packet generation engine must sequentially read all the preceding packets. 

In order to allow quick random access, our delta encoding scheme can be extended to include \emph{entry points}. For example, the FM software layer can store packets $P_{10}, P_{20}, P_{30}, \ldots$ in uncompressed form. Thus, when the software layer instructs the packet generation engine to access the $21^{st}$ packet, it also provides the uncompressed $P_{20}$, and the engine can use $P_{21}$ with one memory access. In this example every packet in the FM memory can be reached in at most $9$ memory access operations. 

\begin{comment}
For $j=1$ to $N$
  Read packet $j$
  Decompress packet $j$
  Transmit packet $j$
  Wait period/N

Remove(k,P)

For $j=1$ to $k-1$
  Read packet $j$
  Decompress packet $j$
  Transmit packet $j$

Decompress packet $P$
Transmit packet $P$
Write packet $P$ to the previous location of packet $k$

For $j=k+2$ to $N$
  Read packet $j$
  Decompress packet $j$
  Transmit packet $j$
  Write packet $j$ to the previous location of packet $j-1$

Insert(k,P,P’)

For $j=1$ to $k-1$
  Read packet $j$
  Decompress packet $j$
  Transmit packet $j$

Decompress packet $P$
Transmit packet $P$
Write packet $P$ to the previous location of packet $k$

Decompress packet $P’$
Transmit packet $P’$
Write packet $P’$ after packet $P$

For $j=k+1$ to $N$
  Read packet $j$
  Decompress packet $j$
  Transmit packet $j$
  Write packet $j$ after previously written packet
\end{comment}

%\vspace{-1mm}
\section{Evaluation}
\label{EvaluationSec}
%\vspace{-2mm}
\subsection{Data Set}
We evaluated our FM-Delta compression system on 100 sets of synthetically generated packets.\footnote{We did not use publicly available packet traces since these traces either do not include FM packets, or do not include packet payloads, thus preventing effective data compression analysis.} Each set comprises 100k packets. Our synthesized data sets are publicly available~\cite{datasets}.

\begin{sloppypar}
The data sets consists of two types of packets: Continuity Check Messages (CCM)~\cite{Y1731}, and Bidirectional Forwarding Detection (BFD)~\cite{RFC5880} control messages. Each packet type was used on half of the data set. CCMs are defined over Ethernet, while BFDs are over IPv4-over-Ethernet. Each packet included a random number of VLAN tags (either 0, 1, or 2 tags).
\end{sloppypar}

The network topology can significantly affect the extent to which FM packets can be compressed in our setting. For example, if multiple FM packets are sent to the same destination device, then some of the packet fields may be similar in these packets. Moreover, in CCMs~\cite{Y1731} the \textbf{MEG ID} field is a 48-byte field that has a different value in each Maintenance Entity Group (MEG)~\cite{Y1731}. Thus, the number of devices per MEG can significantly affect the similarity among packets.

We assumed that the current device has a set of 32 MAC addresses, and thus the source MAC address of each packet was randomly chosen from the pool of 32 addresses, while the destination MAC address was randomly chosen\footnote{Throughout this section the `randomly chosen' refers to a uniformly distributed selection.}  without constraints. The IP addresses of each BFD packet were randomly chosen. The VLAN IDs of each packet that included a VLAN tag was also randomly chosen. The MEG ID field in CCM packets was randomly chosen for each set of 3 CCM packets.\footnote{We assumed that the number of Maintenance Points (MP)~\cite{Y1731} per Maintenance Entity Group (MEG)~\cite{Y1731} is 4 on average.}

Our experiments were performed in two modes: \textbf{ordered} mode, in which packets were arranged so as to allow a higher compression ratio, and \textbf{random} mode, in which packets were ordered randomly. In the ordered mode, the arrangement was implemented according to two criteria: (i) packets were ordered according to their size, allowing alignment between the fields of consecutive packets, and (ii) packets from the same MEG were grouped together.

\subsection{Results}
\begin{figure}[t]
	\centering
  \begin{subfigure}[t]{.24\textwidth}
  \centering
  \fbox{\includegraphics[height=4.8\grafflecm]{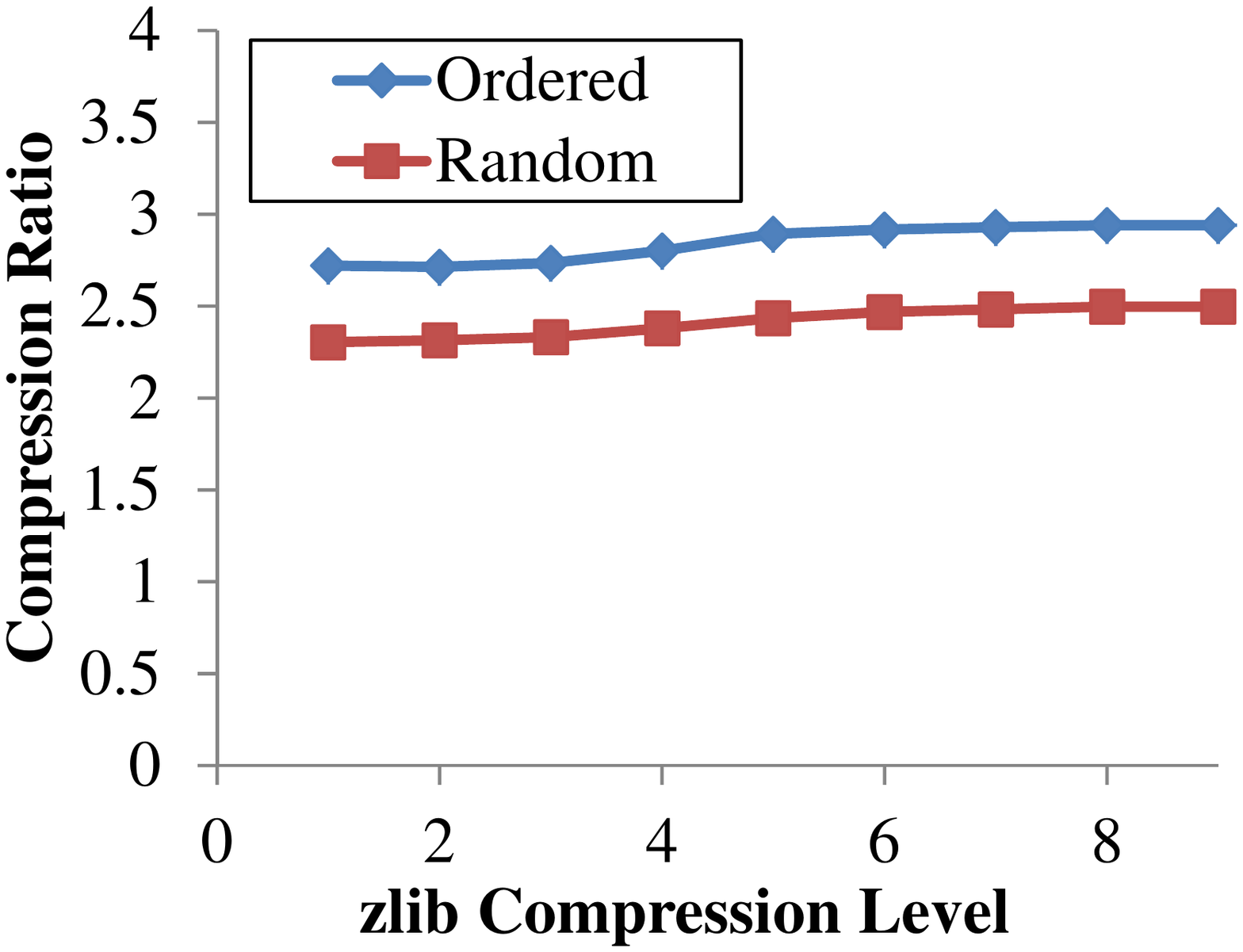}}
	\captionsetup{justification=centering}
  \caption{zlib: Compression ratio as a function of the zlib compression~level.}
  \label{fig:zlib}
  \end{subfigure}%
  \begin{subfigure}[t]{.25\textwidth}
  \centering
  \fbox{\includegraphics[height=4.8\grafflecm]{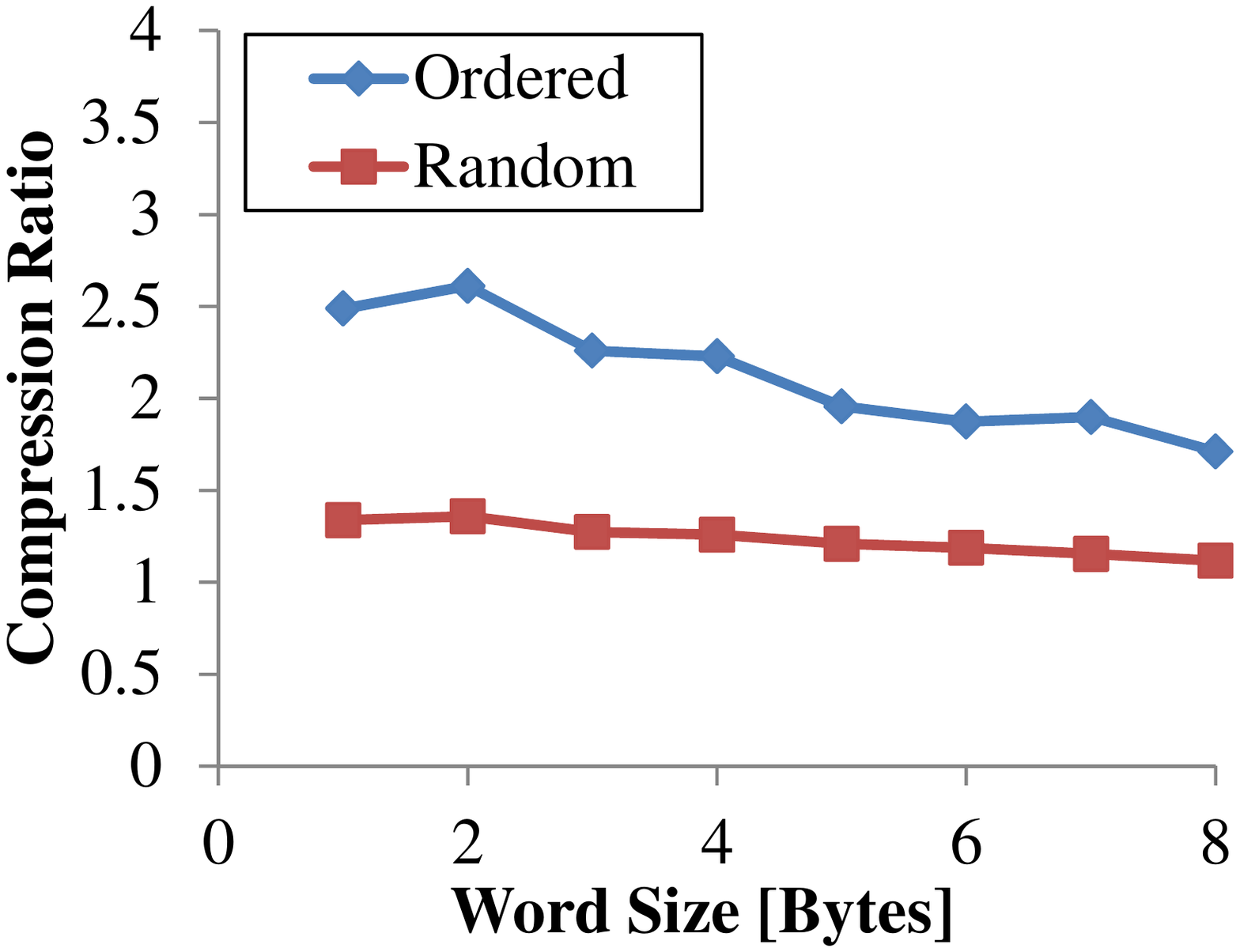}}
	\captionsetup{justification=centering}
  \caption{FM-Delta: Compression ratio as~a~function of the word size.}
  \label{fig:delta}
  \end{subfigure}%
	%\ifdefined\CutSpace\vspace{-3mm}\fi
  \caption{Compression ratio of the two analyzed algorithms. The `Ordered' curve is most interesting, since we have full control over the order of packets in the FM packet memory. In FM-Delta the best compression ratio is achieved when the word size is fixed at 2 bytes.}
  \label{fig:CompRatio}
	%\ifdefined\CutSpace\vspace{-5mm}\fi
\end{figure}

We present a glimpse at some of our experimental results. Fig.~\ref{fig:zlib} illustrates the compression ratio\footnote{The compression ratio is the ratio between the uncompressed data set and the compressed data set.} using zlib. Since zlib supports nine possible compression levels, the graph presents the compression ratio for each of the levels. 

Fig.~\ref{fig:delta} presents the compression ratio of the delta encoding algorithm as a function of the word size. The word size is a design parameter of the FM-Delta algorithm. Specifically, for the FM protocols that were analyzed in this work, if the word size is fixed at 2 bytes, we expect the algorithm to provide the best performance for the FM protocols we analyzed.

Notably, although the delta encoding algorithm is significantly simpler, it provides a comparable compression ratio; zlib provides a compression ratio of 2.9 at the highest compression level, but is not hardware-friendly, while the hardware-friendly FM-Delta provides a ratio of 2.6.

Another significant result is that the ordered message set allows a higher compression ratio, emphasizing the advantage of optimizing the order of the data set. Such an optimization is feasible in light of the relative flexibility of FM specifications as to the order and exact timing of packet transmission.

\ifdefined\CutSpace\vspace{-2mm}\fi
\section{Conclusion and Outlook}
In this paper we have shown that packet compression can significantly reduce the required on-chip memory in Fault Management protocol implementations.
Surprisingly, delta encoding, which is typically dismissed as an ineffective compression approach, is shown to be highly effective for packet compression. We introduce a simple and hardware-friendly delta encoding algorithm that provides a compression ratio of~2.6, and allows to reduce the cost of packet processor silicons. 

Potential improvements of FM-Delta can be considered, for example by using a dictionary for common field values. Furthermore, the preliminary results presented in this paper can be further established by experimenting with a hardware implementation of FM-Delta, and by analyzing real-life traces of FM messages. Notably, the concepts we presented may be applied not only to FM messages, but also to other types of control messages that are stored by switches and routers.
%A natural next step would be to test the algorithm we presented on real-life traces of FM messages from a network operator.
%Sequential read procedure.
%Order.
%future: test the algorithm on real traces

\bibliographystyle{ieeetr}
\bibliography{OAMComp}

\begin{thebibliography}{10}

\bibitem{Fmdelta}
T.~Mizrahi, Y.~Revah, Y.~Refael~Kalim, E.~Kapuza, and Y.~Cassuto, ``{FM-Delta:
  Fault Management Packet Compression},'' in {\em IEEE International Symposium
  on Integrated Network Management}, 2017.

\bibitem{ONFPIF}
``{OF-PI: A Protocol Independent Layer},'' ver. 1.1, {ONF}, 2014.

\bibitem{mizrahi2014overview}
T.~Mizrahi, N.~Sprecher, E.~Bellagamba, and Y.~Weingarten, ``{An Overview of
  Operations, Administration, and Maintenance (OAM) Tools},'' {RFC 7276}, IETF,
  2014.

\bibitem{IEEE8021ag}
{IEEE 802.1ag}, ``{Connectivity Fault Management},'' 2007.

\bibitem{Y1731}
{ITU-T G.8013/Y.1731}, ``{Operations, administration and maintenance (OAM)
  functions and mechanisms for Ethernet-based networks},'' 2015.

\bibitem{RFC5880}
D.~Katz and D.~Ward, ``{Bidirectional Forwarding Detection (BFD)},'' {RFC
  5880}, IETF, 2010.

\bibitem{EricssonSPO1400}
``{Ericsson SPO 1400 Family},'' tech. rep., 2012.

\bibitem{Huawei}
``{Huawei OptiX OSN 550 and OSN 3500},'' tech. rep., 2011.

\bibitem{mizrahi2011oam}
T.~Mizrahi and I.~Yerushalmi, ``The {OAM} jigsaw puzzle,'' technical white
  paper, {Marvell}, 2011.

\bibitem{pica8}
``Pica8 32x40 gbe,''
  \url{http://www.pica8.com/wp-content/uploads/2015/09/pica8-datasheet-32x40gbe-p5401.pdf},
  2014.

\bibitem{rottenstreich2014compressing}
O.~Rottenstreich, M.~Radan, Y.~Cassuto, I.~Keslassy, C.~Arad, T.~Mizrahi,
  Y.~Revah, and A.~Hassidim, ``Compressing forwarding tables for datacenter
  scalability,'' {\em IEEE Journal on Selected Areas in Communications},
  vol.~32, no.~1, pp.~138--151, 2014.

\bibitem{retvari2013compressing}
G.~R{\'e}tv{\'a}ri, J.~Tapolcai, A.~K{\H{o}}r{\"o}si, A.~Majd{\'a}n, and
  Z.~Heszberger, ``Compressing ip forwarding tables: towards entropy bounds and
  beyond,'' in {\em ACM SIGCOMM}, 2013.

\bibitem{tye2003review}
C.~S. Tye and G.~Fairhurst, ``A review of {IP} packet compression techniques,''
  {\em Proc. PGNet}, p.~13, 2003.

\bibitem{gutwin2006improving}
C.~Gutwin, C.~Fedak, M.~Watson, J.~Dyck, and T.~Bell, ``Improving network
  efficiency in real-time groupware with general message compression,'' in {\em
  conference on Computer supported cooperative work}, 2006.

\bibitem{rfc2616}
J.~Mogul, L.~M. Masinter, R.~T. Fielding, J.~Gettys, P.~J. Leach, and
  T.~Berners-Lee, ``{Hypertext Transfer Protocol -- HTTP/1.1}.'' RFC 2616, Mar.
  2013.

\bibitem{websiteOptimization}
``{HTTP Compression},''
  \url{http://www.websiteoptimization.com/speed/tweak/compress/}.

\bibitem{zlib}
J.-l. Gailly and M.~Adler, ``{zlib},'' \url{http://www.zlib.net/}.

\bibitem{datasets}
T.~Mizrahi, Y.~Revah, Y.~Refael~Kalim, E.~Kapuza, and Y.~Cassuto, ``{The
  FM-Delta Project},'' \url{https://sites.google.com/site/fmdeltacompression/}.

\bibitem{ziv1978compression}
J.~Ziv and A.~Lempel, ``Compression of individual sequences via variable-rate
  coding,'' {\em IEEE transactions on Information Theory}, vol.~24, no.~5,
  pp.~530--536, 1978.

\end{thebibliography}

\end{document}